# Neural-network-inferred initial guesses for photoelastic force inversion

Jr-Jiun Lin and Jih-Chiang (JC) Tsai

Institute of Physics, Academia Sinica, Taipei, Taiwan

## Abstract

The inversion of contact forces between closely packed photoelastic disks has become a standard tool in the granular research community. The process iteratively fits the image generated by a forward model to the experimental image, and the correct convergence of the iteration depends critically on the initial guess of the force vectors. In this work, we train a convolutional neural network to infer a force vector from the image in the vicinity of individual contacts. This method, called Contact-Region Inference of Forces, or CRIF, provides the initial guess for the iteration. When tested with synthetic disk images, our approach substantially improves the accuracy of fitting results in comparison to that from the $G^2$ method in the widely adopted PeGS pipeline [1, 2], particularly for disks with many contacts and large forces involving high fringe orders. In addition, we extend the method to disks with a built-in axisymmetric residual stress, which often arises when photoelastic disks are submerged in fluids. Finally, on an experimental packing, fits started from the CRIF guess satisfy force reciprocity at strong contacts where fits started from $G^2$ do not.

## 1. Introduction

Dense granular materials carry load through a heterogeneous network of interparticle contacts, and the statistics and organization of these contact forces underlie their mechanical response[3–6]. In two-dimensional experiments, photoelastic particles make these forces visible: a disk viewed between crossed circular polarizers displays fringes whose pattern encodes the stress field inside it, and therefore the forces acting at its contacts[1, 5, 7]. Photoelasticimetry has become the standard route to particle-scale or contact-scale force measurements in quasi-two-dimensional granular experiments, revealing remarkable properties of these systems such as shear jamming and fabric anisotropy[5, 8–11], the yielding transition[12, 13], and many others[14–17].

Early studies in the field relied primarily on measuring the squared gradient of the transmitted light intensity, commonly referred to as $G^2$, which can be calibrated to serve as a proxy for the pressure at the particle scale[18, 19]. Zhao *et al.* showed that, at the scale of a single disk, $G^2$ is proportional to the sum of the contact-force magnitudes as long as the forces remain below a saturation force[20]. More recent developments have allowed quantitatively accurate recovery of individual contact force vectors, made possible by solving an inverse problem: given the contact positions on a disk, one searches for the set of contact-force vectors whose theoretical fringe pattern best matches the measured image, subject to force and torque balance[1, 5]. Mature open-source tools now exist for extracting forces from images, such as the popular PeGS algorithm [1, 7].

This fitting approach is a nonlinear least-squares problem whose objective is non-convex, and the solution it reaches is a local minimum that depends strongly on where the optimization starts[1]. The quality of the initial guess is thus key for this method to yield accurate measurements. In PeGS2[2], for example, the $G^2$ of each particle is first converted into the total force $f_{tot}$ on the particle through a calibration. $f_{tot}$ is then shared among the contacts in proportion to the $G^2$ measured in a local region at the contact point, and all contact forces are initially taken to be purely normal[1, 2]. This works best when the fringes of different contacts are nearly independent, i.e., for small forces and few contacts, and breaks down otherwise. For 6-mm-thick ClearFlex50 disks, the reliability of the standard PeGS implementation was found to drop considerably when forces exceed 1.2–2.5 N, depending on the number of contacts[14]. This severely restricts the range of experimental conditions in which quantitative force inversion is possible.

On the other hand, recent advances in machine learning and computer vision offer an alternative way to extract forces from photoelastic images. Lin[21] proposed an artificial neural network that can infer the total force from $G^2$ and other known parameters, bypassing the force calibration. Sergazinov *et al.* [22] trained convolutional neural networks (CNNs) to predict contact number, contact positions, force magnitudes, and contact angles from synthetic images of individual particles. These innovations highlight the remarkable potential that neural networks can bring to photoelasticimetry. However, these approaches are largely separate from the widely adopted PeGS framework, whose physics-based fit has proven reliable in the literature.

In this paper, we develop and test an integrated approach that combines the strengths of the CNN inference and the physics-based PeGS fit. Because contact numbers and positions can be determined accurately from particle tracking and simple geometric relations, we take them as known, which simplifies the problem. We train a ResNet-18 network [23] on synthetic images of individual *contact regions*, not particles, to infer the force vectors as the starting values for the iterative process. We refer to this approach as Contact-Region Inference of Forces (CRIF). We demonstrate with synthetic images that the CRIF guess substantially reduces the error of the final fit when compared to the $G^2$ method, and remains robust for fringe orders up to six. In addition, we extend this method to disks with halos, which are present when disks are immersed in fluid[24] or manufactured by casting. We propose a residual-stress model that reproduces the observed halo, show that accurate forces are recovered once the halo is included in the forward model, and demonstrate on images of a sheared packing that fits started from CRIF remain robust while fits started from the $G^2$ guess fail. The code and trained networks are publicly available[25].

## 2. Existing methods for photoelastic force inversion

In this section we provide a brief overview of the principles of photoelasticity, followed by the PeGS force inversion technique and its limitations. The topics covered in this section are discussed in greater detail in Refs.[1, 26].

### *2.1 Photoelasticity*

We consider a piece of photoelastic material with uniform thickness viewed under a dark-field circular polariscope. The transmitted intensity field is given by the stress-optic law [27]:

$$I(x,y) = I_0 \sin^2\left(\frac{\pi\Delta\sigma(x,y)h}{\lambda/C}\right) \qquad Eq.1$$

, where $\Delta\sigma = \sqrt{(\sigma_{xx} - \sigma_{yy})^2 + 4\sigma_{xy}^2}$ is the local difference between the two in-plane principal stresses. $I_0$ is the illumination intensity, $\lambda$ is the wavelength of light, and $C(\lambda)$ is the wavelength-dependent stress-optic coefficient of the material. The photoelastic coefficient $F_\sigma \equiv \lambda/Ch$ sets the stress scale of the photoelastic response. Another quantity that often arises in discussing photoelasticity is the fringe order, as demonstrated by a diametrically loaded disk in Fig. 8(a) of Ref.[1]. Here we define the fringe order of a contact force of magnitude $f$ to be $N_{fr}(f) = \frac{4f}{\pi R F_\sigma h}$, consistent with Ref.[1]. This is equivalent to the number of fringes counted from the rim to the center if the disk were loaded diametrically by two opposing forces $f$. The value of $N_{fr}$ is thus a convenient measure of the fringe density near a contact of force $f$ that can be compared across disk sizes and materials.

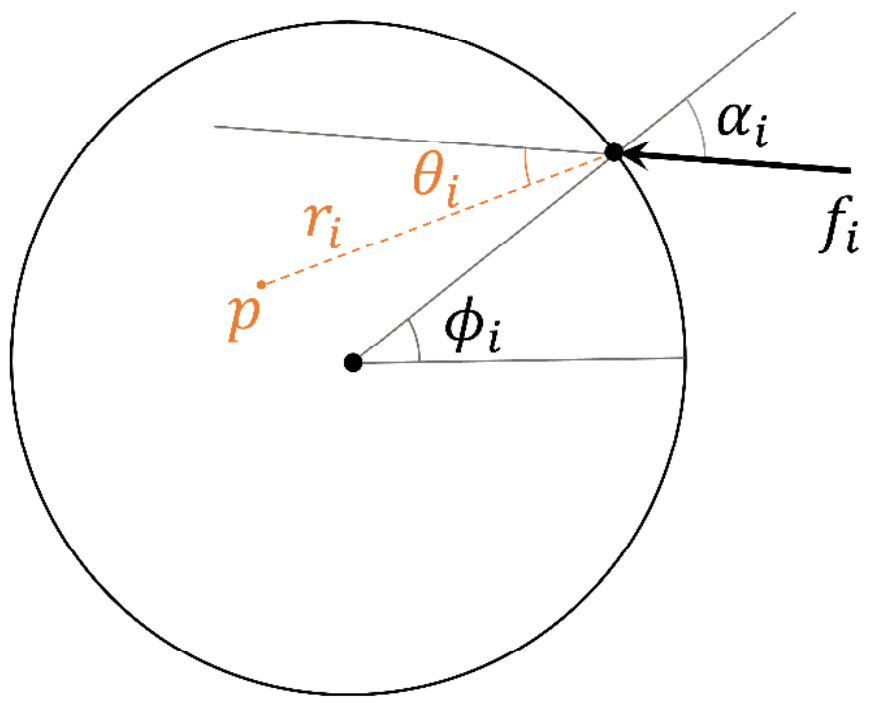

Fig. 1 A force of magnitude $f_i$ is applied with an angle of attack $\alpha_i$ on the circumference of a disk, at a position specified by $\phi_i$. Any point $p$ inside the disk can be defined by the polar coordinates $(r_i, \theta_i)$ with respect to the direction of the force. All angles are measured counterclockwise; for clarity, $\alpha_i$ is drawn with a negative value.

In granular experiments, each disk is loaded through discrete point contacts. The $i$th contact is described by its position angle $\phi_i$ on the circumference, the magnitude $f_i$ of the contact force, and the angle $\alpha_i$ between the force and the inward normal (Fig. 1). Each contact contributes a stress field $\boldsymbol{\sigma}_i^0$ which is purely radial in the polar coordinates $(r_i, \theta_i)$ centered at the contact and measured from the force direction (Appendix A). The total contact stress $\boldsymbol{\sigma}_c$ over $Z$ contacts is [26]:

$$\boldsymbol{\sigma}_c(x, y) = \sum_{i=1}^{Z} \boldsymbol{T}(\gamma_i)\boldsymbol{\sigma}_i^0\boldsymbol{T}^T(\gamma_i) + \sum_{i=1}^{Z} \frac{f_i \cos(\alpha_i)}{2\pi R}\boldsymbol{I} \qquad Eq.\mathbf{2}$$

, where $\boldsymbol{T}(\gamma_i)$ rotates $\boldsymbol{\sigma}_i^0$ by an angle $\gamma_i = \theta_i - \alpha_i - \phi_i$. The second term imposes a stress-free boundary at $r = R$; it is isotropic and therefore does not enter $\Delta\sigma$. Given a set of force vectors on the boundary, $\{\phi_i, f_i, \alpha_i\}$, Eqs. 1 and 2 thus give the image analytically, a mapping referred to herein as the forward model. The inverse process of obtaining $\{\phi_i, f_i, \alpha_i\}$ from an experimental image is considerably more challenging, and is the task the PeGS codebase [2] is built to solve.

### *2.2 PeGS pipeline*

The Photoelastic Grain Solver (PeGS) is an open-source implementation of the force-inversion method of Majmudar and Behringer [5], written by J. E. Kollmer and introduced by Daniels *et al.* [1]. Its current version, PeGS2 [2], adopts the corrected treatment of the stress-free boundary term. For each image, PeGS2 locates the particles, identifies contacts, and then fits the contact forces on each particle by nonlinear least squares, as described below.

Particle positions and the list of contacting pairs are taken as known here. Particle centers can be found with circle detection algorithms [2] or more recent CNN-based tools [28, 29], and force-bearing contacts can be identified with a $G^2$ threshold [1] or from the intensity profile along the circumference [10].

For each disk, the knowledge of particle positions and contact pairs determines the contact angles $\phi_i$, which leaves $\{f_i, \alpha_i\}$ as the unknowns. The fitting process starts with an initial guess, $\{f_i, \alpha_i\}^0$, and generates the corresponding image $I(\{f_i, \alpha_i\}^0)$ via the forward model. The program then iteratively tunes $\{f_i, \alpha_i\}$ to minimize the squared pixel-wise difference between the forward model prediction and the experimental image [5].

The periodic nature of $I(\boldsymbol{\sigma})$ in Eq. 1 implies that the inverse problem is highly non-convex, and the fitting can only converge to a nearby local minimum. The choice of initial guess therefore determines the accuracy of the measured forces, not only the speed of the fit. The current PeGS2 implementation uses a $G^2$-based initial guess, as did the earlier version described in Ref.[1]. The approach is described below.

### *2.3 $G^2$ initial guess*

The current PeGS2 implementation [2] provides an initial guess based on the local $G^2$ magnitude. The squared gradient over the entire particle area $G_p^2$ can be evaluated with the central difference, as in Eq. 1 of Ref.[1]. It is converted to a total force through a linear relation, $f_{tot} = \Gamma G_p^2$. $\Gamma$ is an empirical factor that is usually calibrated experimentally by loading a single disk, and $f_{tot}$ is the sum of force magnitudes over all contacts. This linear relation has been verified to be robust across various loading conditions and contact numbers [20]. The initial guess of individual contact force magnitudes $\{f_i^0\}$ is given by distributing $f_{tot}$ among the contacts in proportion to the squared gradient measured in a small circular region next to the $i$th contact, namely:

$$f_i^0 = f_{tot} \frac{G_i^2}{\Sigma_j G_j^2} \qquad Eq.3$$

, $G_j^2$ is the local $G^2$ in the vicinity of the $j$th contact, usually taken as the circular region tangent to the circumference of the disk at the contact point, with a radius $r_c$ typically less than half of the disk radius. All contact forces start as purely normal: $\alpha_i^0 = 0$. PeGS2 then applies a force-balance correction before fitting.

The method above becomes unreliable at higher $N_{fr}$ and $Z$ for two main reasons. First, apportioning $f_{tot}$ by $G_i^2$ is heuristic: the fringes near one contact are also shaped by the other contacts, and near the contact points they are concentrated and blurred, so that $G_i^2$ no longer measures the force at that contact. Second, the guess contains no information about the force directions, which the force-balance step can fix only in simple cases such as $Z = 2$. Both problems worsen with increasing $N_{fr}$ and $Z$, i.e., in densely packed frictional systems. These limitations have been quantified in the appendix of Ref.[14] and are further characterized in Sec. 4.1.

## 3. Our approach, CRIF

To improve the accuracy of the initial guess, we train a CNN to infer force vectors from the image of individual contact regions. The training data are synthetic images generated by the forward model, as is common in past research [21, 22]. In this section, we describe the core methods: the sampling and generation of these synthetic images, the model architecture and training procedure, and the PeGS-style fitting algorithm that we adopt. Further implementation details are given in Appendices B and C.

### *3.1 Synthetic image generation*

Each synthetic image shows one disk of radius 6 mm, with $Z = 2\sim6$ contacts. We set $f_\sigma \equiv F_\sigma h = 225\ \mathrm{N/m}$ to resemble the popular photoelastic material, ClearFlex50, with thickness $h = 10\ \mathrm{mm}$. The force sets $\{\phi_i, f_i, \alpha_i\}$ are drawn at random, with a minimum angular separation between $\phi_i$ being $53°$ and a total force between 0.02 and 4 N, and are then corrected to satisfy force and torque balance. Sets with any force above $N_{fr} = 6$ are rejected to set a well-defined upper bound (see Appendix B for the full procedure). The resulting forces span 0.01–6.4 N, with nearly uniform coverage in $\ln(f)$, and force angles up to 45° [Fig. 2(a, b)].

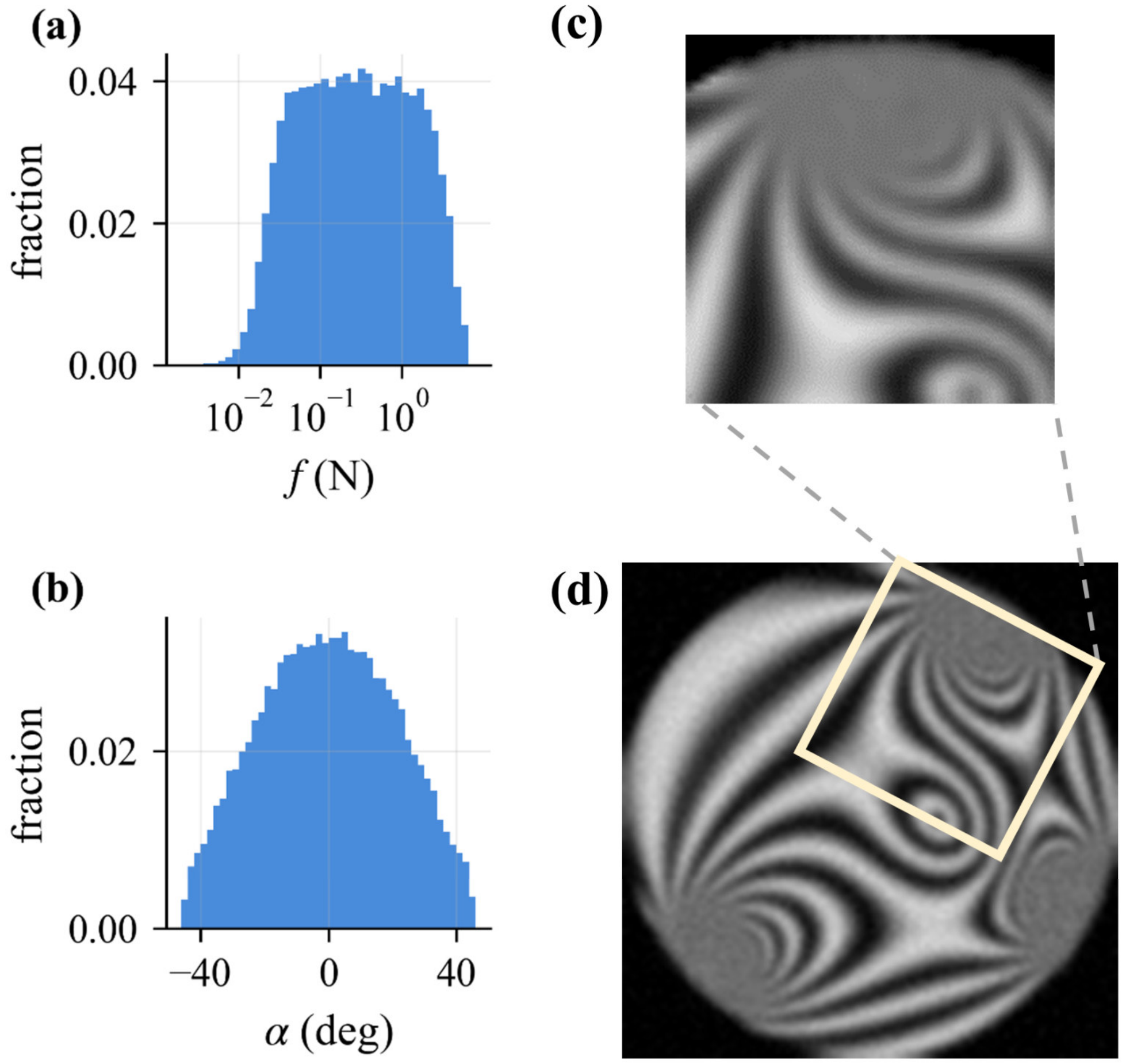

Fig. 2 The distribution of (a) force magnitude $f$ and (b) angle $\alpha$ in the force sets generated with the method described in Sec. 3.1 and Appendix B. (c) Demonstration of the contact-region image crop. A 0.92R × 0.92R square is cropped such that the contact point lies at the top center of the cropped square, and the contact normal points in the vertical direction. The disk image (d) is rendered at a diameter of 100 pixels, with Gaussian blur $\sigma = 1$ pixel and Gaussian noise $\sigma = 0.015$.

Each disk is rendered at a diameter of 100 pixels and undergoes a series of modifications with random gain $g$, offset $b$, Gaussian blur, additive Gaussian noise, and quantization to 8 bits (Appendix B). These modifications mimic experimental imperfections and help the model generalize. We generate 3,000 force sets for each $Z$. The 15,000 sets are randomly split 70/15/15 into training, validation and test sets, and each set is rendered at two random modification settings, giving 21,000 training, 4,500 validation and 4,500 test images of disk.

### *3.2 Training dataset*

The image sets described above are similar to those used in Ref.[22] to train four separate models that predict $Z$, $\phi$, $f$, and $\alpha$. This design makes training unnecessarily difficult. First, $Z$ and $\phi$ can be measured accurately with the techniques mentioned in Sec. 2.2 and treated as known inputs. Second, the result should be rotationally symmetric; however, the same $\{f_i, \alpha_i\}$ but rotated can produce images that look very different to the convolution layers. Third, a network outputs a fixed number of values, while the number of contacts varies from disk to disk--- Ref.[22] therefore trained a model for each value of $Z$, *i.e.* 15 models in total.

We propose a simpler approach that largely overcomes these challenges: Train with images of *contacts*, not disks. For each contact in a disk image, we crop out a $0.92R \times 0.92R$ square region such that the contact point lies at the top center of the crop, and the contact normal aligns with the vertical, as demonstrated in Fig. 2(c). The process is repeated for each $i$, generating $Z$ contact images from one disk image. This removes the arbitrary orientation of the disk and the $Z$ uncertainty, yielding 84,000 training, 18,000 validation, and 18,000 test contact-region images that are all aligned in the same direction and contain information about the local fringe pattern.

### *3.3 Model training*

We use a ResNet-18 architecture[23] pretrained on ImageNet, with its classifier replaced by a small regression head with two outputs (Appendix C). Training converges overnight on a single consumer GPU (NVIDIA GTX 1060).

### *3.4 Force fitting*

In order to characterize how the quality of the guess influences the fitting result, we set up a PeGS-style fitting process in Python, written in PyTorch so that many disks are fitted in parallel on a GPU. The forward model of Eq. 2 generates the synthetic intensity map at the resolution of the image, which is then smoothed with a 3 × 3 Gaussian kernel $S$, with a standard deviation of one pixel, to create the forward image $I^*(\{f_i, \alpha_i\})$. We note that the random gain and offset (Sec.3.1) are meant to help the CRIF network generalize to various lighting conditions; in the fitting process they are removed exactly to create the clean target $T = (S * I - b)/g$, which can then be compared with $I^*$ without an explicit contribution from $b$ and $g$. The objective of the optimizer is then to minimize the loss

$$L = [I^*(\{f_i, \alpha_i\}) - T]^2 \qquad Eq. 4$$

, by updating $\{f_i, \alpha_i\}$ with the Adam optimizer at a learning rate of 0.04.

We apply two stopping rules (Table 1): a loose fit which stops once $L$ has failed to decrease by $10^{-3}$ for 10 consecutive iterations, and a tight fit that has much longer patience and allows more precise convergence. The tight fit is much more expensive, and the loose fit is often implemented in practice. We present results from both settings for completeness.

**TABLE 1.** Settings of the force fit.

| | **Loose fit** | **Tight fit** |
|---|---|---|
| tolerance on the decrease of $L$ | $10^{-3}$ | $10^{-6}$ |
| patience (iterations) | 10 | 100 |
| maximum iterations | 1000 | 2000 |

## 4. Comparison of G-squared and CRIF results

### *4.1 $G^2$ guess*

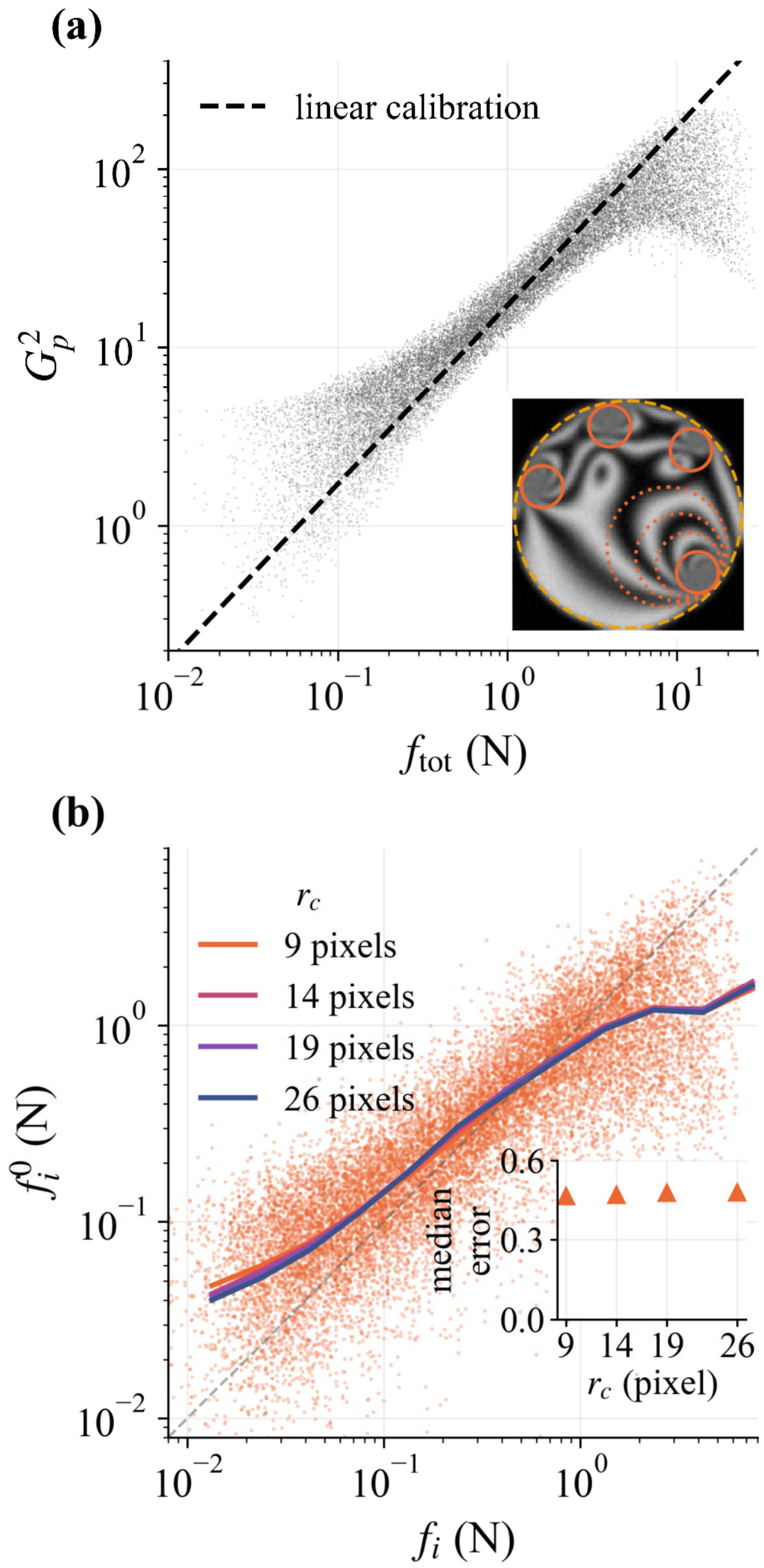

Fig. 3 The gradient-squared guess on disks. (a) $G_p^2$ against the total contact force $f_{tot}$ on the 21,000 training disks, one dot per disk, with the linear calibration $f_{tot} = \Gamma G_p^2$ (dashed). Inset: a sample schematic of the $G^2$ method. The large dashed circle marks the disk area over which $G_p^2$ is evaluated. Solid orange circles at each contact show the $r_c = 9$ pixel contact regions where $G_i^2$ is evaluated; the dotted circles show the various $r_c$ values used. (b) Guessed $f_i^0$ against true force $f_i$ for each contact in the test set. Solid lines of different colors show the bin medians (see main text for its definition) for the four $r_c$ values. Scattered dots are individual guesses using $r_c = 9$ pixels. The dashed diagonal is a guide to the eye. Inset: Median of the relative error $|f_i^0 - f_i|/f_i$ over all contacts in the test set against $r_c$.

We first examine the $G^2$ guess on synthetic images, since it is the standard starting point for the fit and sets the baseline against which CRIF is judged. Because the squared gradient scales with the square of the gain, all tests concerning the $G^2$ guess use images of constant gain $g = 0.8$ and offset $b = 0.025$ for simplicity. We carry out the $G^2$ method in the PeGS pipeline as described in Sec. 2.3, where we find $f_{tot}$ on the disk from a $G_p^2$ calibration, and distribute the forces to each contact such that $f_i$ is proportional to $G_i^2$, the local $G^2$ within a small circular region of radius $r_c$ near the $i$-th contact point, as demonstrated by the solid orange circles in the inset of Fig. 3(a). The force-balancing step of PeGS2 is also applied.

The first step is to find the calibration relation between $G_p^2$ and $f_{tot}$. We plot these two values for all 21,000 images in the training set as scattered dots in Fig. 3(a), and the linear calibration $f_{tot} = \Gamma G_p^2$ as the dashed line. The calibration constant $\Gamma$ specific to this resolution is fitted to be 0.0586 N, which will be used throughout this paper for all $G^2$ guesses. In practice, this calibration should be done with physical disks in imaging conditions identical to the actual experiment, so it is often the case that only $Z = 2$ and $\alpha = 0$ images are taken as data points due to experimental limitations. This would likely result in a cleaner linear fit over a larger force range. We do not impose this limitation here for the purpose of generality and include all images in the training set.

Using the calibration obtained above, we proceed and make a guess on each contact force in the test set based on Eq. 3. A free parameter in this process is the contact region radius $r_c$. It is not obvious what the optimal $r_c$ should be, so we tested four $r_c$ values from 9 to 26 pixels (shown as dotted circles in Fig. 3(a)). Figure 3(b) shows the guessed $f_i^0$ versus the true $f_i$ for all contacts. Here and in the following figures, we plot bin medians: the horizontal axis is divided into logarithmic bins of 0.25 decade, and the median of the plotted quantity is taken over all points in each bin (bins with fewer than 20 points are omitted). The bin medians for the four $r_c$ values nearly collapse onto one another, and the global median of the relative error $|f_i^0 - f_i|/f_i$ is likewise insensitive to $r_c$, staying above 45%. The guesses $f_i^0$ spread over a decade across the entire range of $f_i$, and the S-shaped bin medians show the same overestimation at small forces and underestimation at large forces as the $G_p^2$calibration.

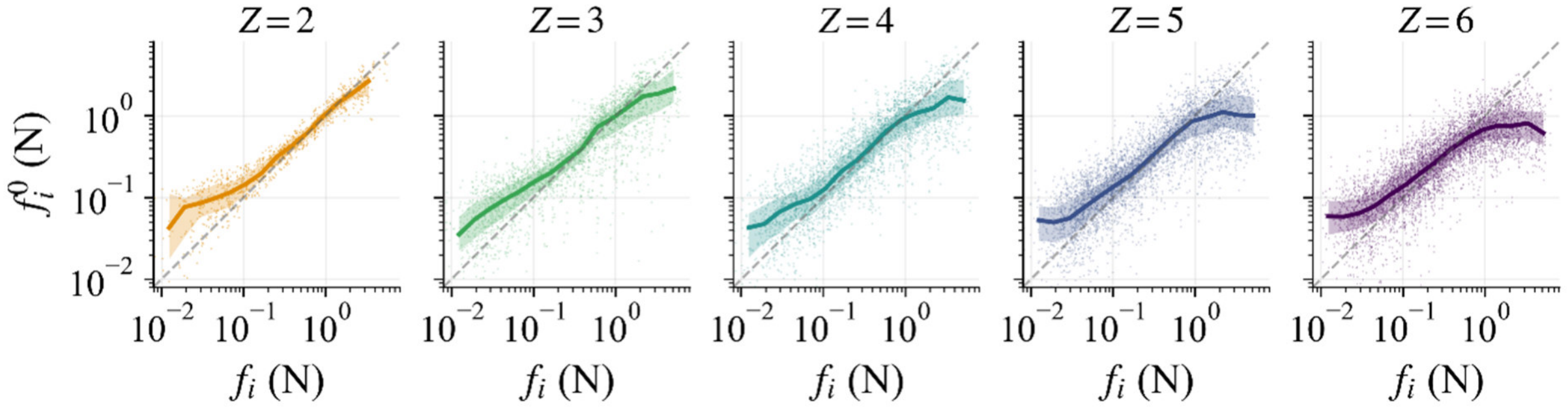

Fig. 4 The $G^2$ initial guess against the true contact force $f_i$ in the test set, distinguished by number of contacts $Z$ (from 1,800 contacts at $Z = 2$ to 5,400 at $Z = 6$). One dot represents one contact. Solid lines represent bin medians. Shaded bands show the spread from the 25th to 75th percentile in each bin.

The considerable error of $f_i^0$ in Fig. 3(b) can be further examined by categorizing contacts based on the contact number $Z$ of the disk they belong to. Figure 4 shows a clear trend of increasing error at larger $Z$. For $Z = 2$, the guess closely follows the true value for $f_i > 0.2$ N, with a 19% spread. The error roughly doubles from Z = 2 to Z = 3, and from there the strong contacts are increasingly underestimated. The force level at which $f_i^0$ begins to flatten decreases with larger $Z$, from $f_i \approx 2N$ at $Z = 3$ down to $f_i \approx 0.5$ N at $Z = 6$, qualitatively similar to what was observed in Ref.[14]. Figures 3 and 4 thus confirm quantitatively that the $G^2$ guess fails when $f_{tot}$ exceeds a few newtons or $Z \geq 3$.

### *4.2 CRIF guess and its comparison to $G^2$ guess*

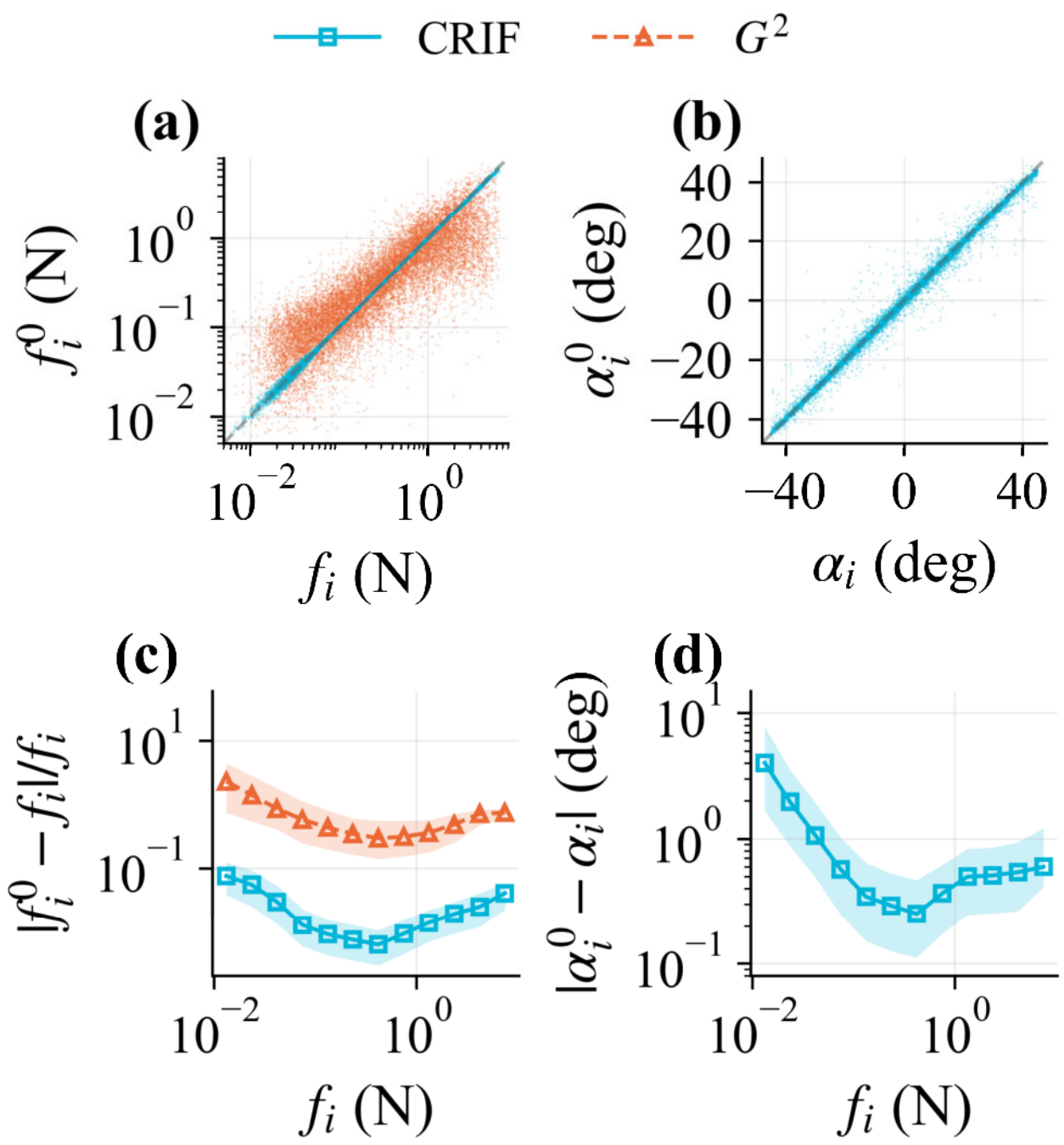

Fig. 5 Initial guesses from CRIF and $G^2$ methods on the 18,000 contacts in the test set. (a) Predicted force magnitude $f_i^0$ and (b) predicted angle $\alpha_i^0$ against the true values, one dot per contact. The dashed line is the diagonal. (c) Relative force error and (d) angle error against the true force; lines show bin medians, and bands show 25th to 75th percentile in each bin. The $G^2$ guess does not predict $\alpha$ and is omitted from (b) and (d).

We now turn to the CRIF guess on the same test contacts. Figure 5(a) shows that the predicted forces follow the true ones over the three decades of the test set, with an overall median of 1.2% in relative error. Unlike the $G^2$ guess, the network also predicts the direction $\alpha$: the median angle error is 0.47°, over angles that span ±45° [Fig. 5(b)]. Figure 5(c, d) shows how the errors in $f$ and $\alpha$ depend on the true force magnitude $f_i$. CRIF performs best for $f_i$ between 0.1 and 1 N, where the bin median error is 0.8% in magnitude and 0.3° in angle. The errors grow toward both ends of the range, likely for different reasons. The weakest contacts produce fringes whose intensity barely exceeds the camera noise, while the strongest contacts produce fringes that approach the pixel scale near the contact point. In addition, the CRIF guess shows no dependence on $Z$. The overall median error ranges from 1.1% to 1.3% in magnitude and 0.43° to 0.50° in angle for $Z = 2\sim6$ (not shown).

The contrast between the two guesses in Fig. 5(c) is large across the whole range. Even in its best range, 0.4–0.6 N, the $G^2$ guess has a bin median error of 29%, about 40 to 50 times that of CRIF. It rises to 74% at 5 N and exceeds 100% below 0.03 N. In contrast, the errors of the CRIF guess stay well below 5% for forces ranging from 0.05 to 5 N. Errors from the PeGS pipeline in high-quality experimental calibrations have been reported to be as low as a few percent, depending primarily on the force magnitudes[14, 26]. We find that CRIF can achieve a similar level of accuracy on unseen synthetic images before any fit. We test in the next subsection whether a good initial guess from CRIF can improve the results of a PeGS-style fitting process.

### *4.3 Comparison after the iterative fit*

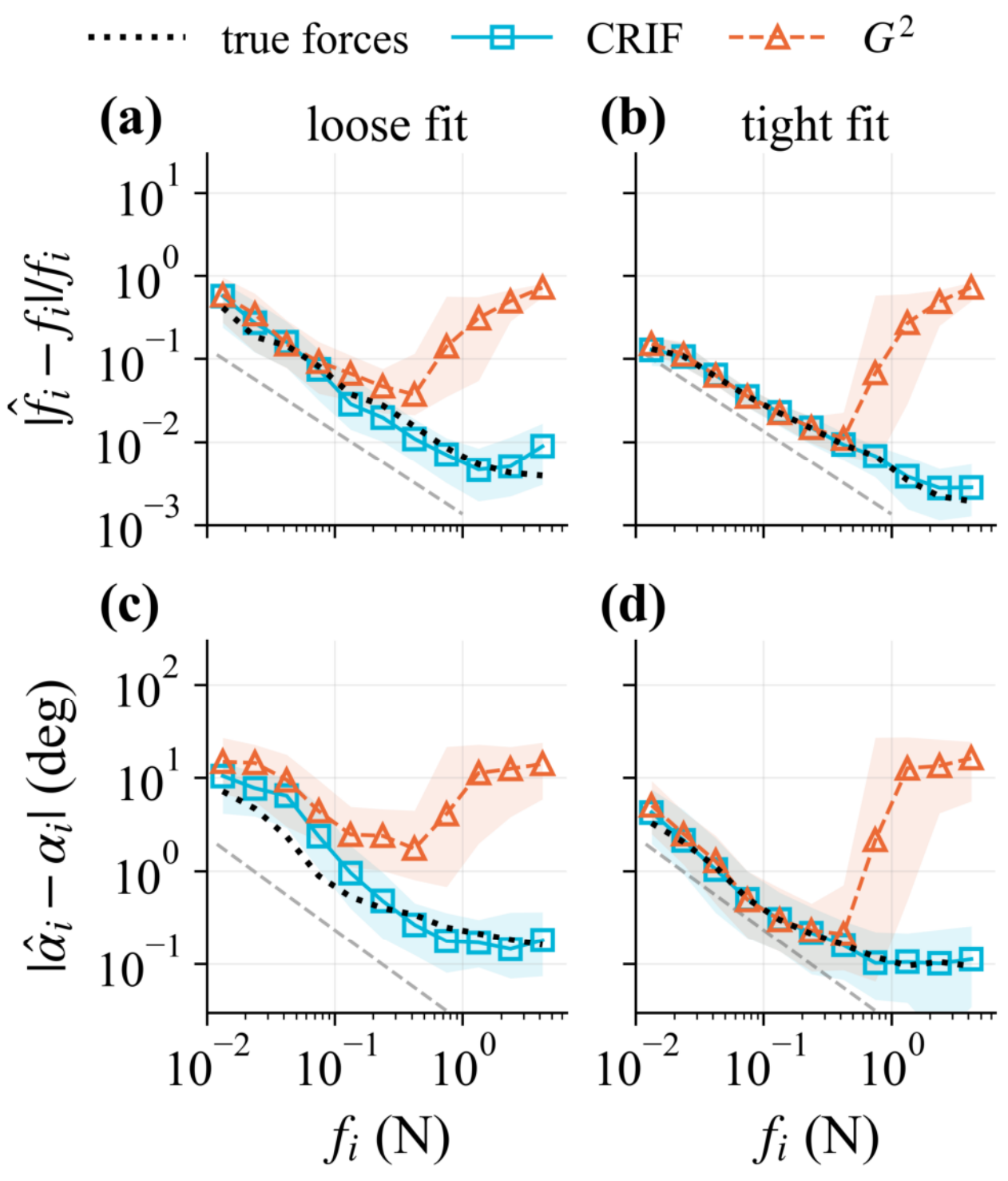

Fig. 6 The fitted force $\hat{f}_i$ and angle $\hat{\alpha}_i$ on 500 synthetic disk images in the test set, started from the true forces, the CRIF guess and the $G^2$ guess. The top row shows the bin median of relative force error $|\hat{f}_i - f_i|/f_i$ after the (a) loose fit and (b) tight fit. The bottom row shows the error in angle $|\hat{\alpha}_i - \alpha_i|$ after the (c) loose fit and (d) tight fit. Lines show bin medians, and bands show the spread from the 25th to 75th percentile, shown for the CRIF and $G^2$ starts. Dashed lines are $\propto f_i^{-1}$ as a guide to the eye.

We fit the test disks using the method detailed in Sec. 3.4 to obtain the fitted force $\hat{f}_i$ and angle $\hat{\alpha}_i$ using various initial guesses. Figure 6 shows the median of the final fitting errors from a loose fit and a tight fit, as defined in Table 1, of 2,000 contacts from 500 randomly sampled disks (100 disks per $Z$) in the test set. Not all disks were tested due to the computational cost of the tight fit. Three different initial guesses are applied for comparison: $G^2$, CRIF, and the true-force guess. For the true-force guess, the fit starts from the true forces; it does not return them exactly because of the modifications applied to the test images. The fitting errors from the true-force guess thus represent an accuracy floor of our test protocol. Figures 6(a) and 6(b) show the relative error $|\hat{f}_i - f_i|/f_i$ of the true-force guess to scale roughly with $f_i^{-1}$ for $f_i < 1$ N, indicating a constant absolute error $|\hat{f}_i - f_i| \approx 10^{-2}$ N for the loose fit, and $2 \times 10^{-3}$ N for the tight fit in the small-force regime. Beyond $f_i > 1$N, the relative error in force flattens out at 0.5% and 0.2% for the loose and tight fit, respectively.

Likewise, the error in $\alpha$ scales roughly with $f_i^{-1}$ in a similar fashion, reaching a constant level of around 0.2 and 0.1 degrees for the loose and tight fit when $f_i > 1$ N.

In all cases, fitting errors from the CRIF guess are comparable to the true-force guess, showing that the CRIF guess is close to optimal in the entire range considered. This is not surprising given the excellent accuracy of the CRIF guess even before the fitting. Interestingly, the error *increases* after the fit for $f_i < 1$ N. This is presumably because the CRIF network, trained on the images modified by added noise and blurring, overcomes these imperfections at the initial guess, whereas the PeGS-style fit that follows tends to converge to a minimum that is shifted by the modification.

On the other hand, the $G^2$ guess underperforms particularly in the large-force regime. For the loose fit, relative errors in both forces and angles reach their minimum at around 0.4 N and rise for larger $f_i$. For $f_i > 1$ N, the median error reaches 50% in force and 10 degrees in angle. The tight fit strategy provides little improvement in this regime, as the optimizer gets stuck in a wrong local minimum. Nevertheless, we notice that the $G^2$ guess performs as well as the other two guesses when $f_i < 0.4$ N despite the larger initial error in Fig. 5(c), showcasing the optimizer's capability at low fringe order.

Thus far, we have characterized the accuracy of the CRIF guess and the resulting fit on synthetic test images, which is the best-case scenario for this approach. Before testing the method on experimental images, we extend the formulation by introducing a residual stress in the stress field, an unwanted feature in photoelasticimetry. We demonstrate how CRIF can be trained to predict forces in the presence of a residual stress term, and apply it to real disks.

## 5. Residual-stress halo and experimental test

In experiments, “halos” at the edge of individual disks, caused by residual stresses, are an undesired feature that interferes with the photoelastic force inversion. In this section, we demonstrate how the CRIF method can accommodate such imperfections by including a modeled residual stress in the forward model. We show that including this term in the fit helps recover accurate forces, while retraining the network on synthetic images with halos speeds up convergence.

### *5.1 Modeling the residual-stress halo*

Both the $G^2$ calibration and the PeGS framework rely on the material being subject only to contact stresses in order to make accurate measurements. This requirement is often not met in practice. Photoelastic disks are not commercially available and are mostly manufactured by casting urethane in a mold, or cutting[1, 7], which can sometimes leave residual stresses that often take the form of a bright halo when viewed under a polariscope, thus interfering with the photoelastic measurement. These residual stresses can stem from the manufacturing process[7] or absorption of ambient moisture during storage[30]. Careful trial-and-error is thus required to optimize the curing protocol and storage conditions in order to minimize residual stresses. In other cases, polyurethane grains that are immersed in aqueous fluids can swell due to absorption. Hayes *et al.*[24] characterized the swelling and photoelastic response of water-clear 55 (WC-55) polyurethane disks immersed in water, showing that the halo forms and fades away over the course of a few months as water permeates the entire disk. On the other hand, we find that our ClearFlex50 disks show a persistent halo after more than a year of immersion in 70% glycerol aqueous solution, as shown in Fig. 7. This material-dependent halo is thus an important factor to be accounted for in wet photoelastic experiments, and has largely restricted accurate force inversion to dry experiments.

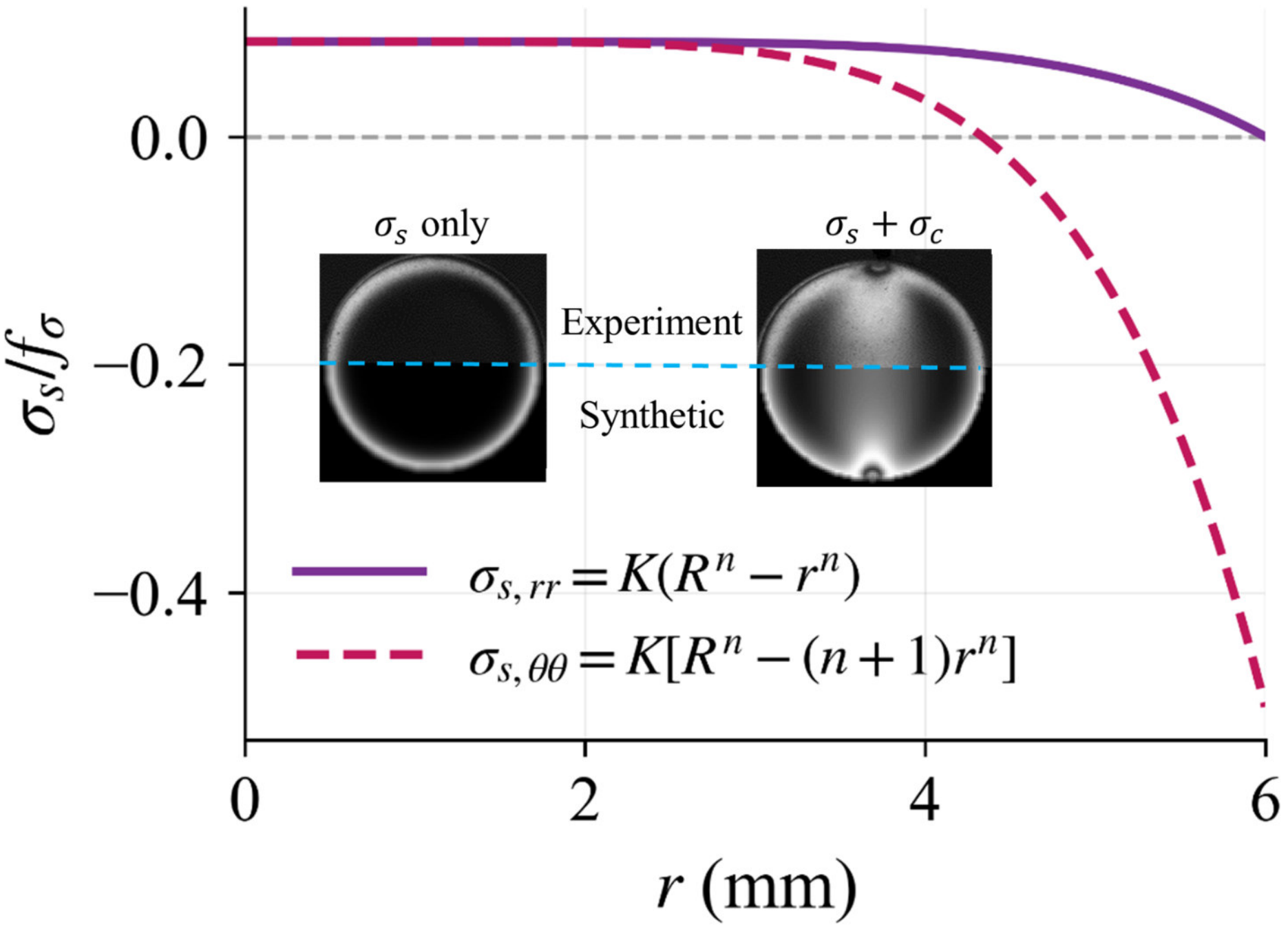

Fig. 7 The radial profile of the two principal components in the modeled residual stress $\boldsymbol{\sigma_s}$. Inset shows the comparison between experimentally captured images (top half) and synthetic images (bottom half) generated using the proposed forward model $\boldsymbol{\sigma_H} = \boldsymbol{\sigma_c} + \boldsymbol{\sigma_s}$. The left panel shows a disk without contacts, and the right panel shows a disk loaded with 0.3 N at the two poles.

Regardless of its origin, the swelling halo poses a major challenge to photoelastic measurements in general, especially for small forces, whose photoelastic response lies mostly near the disk edge. There is an "edge removal percentage" parameter in the current version of the PeGS code where one can ignore the outer rim of a certain width, effectively removing the halo in the fitting. This quick option was shown to yield accurate results for forces larger than 0.3 N in the case of Ref.[24], but is not expected to resolve forces smaller than that.

This limitation can be alleviated if the halo is included in the forward model, both for generating the training images and in the fit. We thus propose to construct an alternative forward model that adds a swelling stress $\boldsymbol{\sigma_s}$ on top of the contact stress $\boldsymbol{\sigma_c}$ in Eq. 2. Namely, we use the hybrid stress $\boldsymbol{\sigma_H} = \boldsymbol{\sigma_c} + \boldsymbol{\sigma_s}$ as the forward model in generating image data as well as in the fitting process. The explicit form of $\boldsymbol{\sigma_s}$ is designed to be axisymmetric, in equilibrium on its own, free of traction at the edge, and concentrated near the rim. In polar coordinates $(r, \theta)$ centered on the disk, we assume:

$$\begin{cases} \sigma_{s,rr} = K(R^n - r^n) \\ \sigma_{s,\theta\theta} = K[R^n - (n+1)r^n] \end{cases} \qquad Eq.5$$

that satisfies $\partial_r \sigma_{s,rr} + \frac{\sigma_{s,rr} - \sigma_{s,\theta\theta}}{r} = 0$, *i.e.* $\nabla \cdot \boldsymbol{\sigma_s} = 0$ with a symmetry over $\theta$, and $\sigma_{s,rr}(r = R) = 0$. $K$ is a scaling constant that can be adjusted based on the observed halo. For our immersed disks, we set $K = f_\sigma / 2nR^n$ such that the image intensity reaches a maximum at $r = R$. The only free parameter left is $n$ that dictates the width of the halo. We use $n = 6$ in this paper since it gives reasonable agreement with the real disks, as shown in the comparison in Fig. 7. In practice, the precise values of $K$ and $n$ can be calibrated in experimental conditions once the halo reaches its steady state.

We introduce the assumed analytical form of $\boldsymbol{\sigma_H}$ into the halo forward model to generate synthetic images from the same force sets used in Sec. 3.1. These data are then used to train an alternative halo-CRIF network using the same recipe as Appendix C. The halo forward model is also incorporated into the fitting process introduced in Sec. 3.4.

### *5.2 Performance of halo-CRIF*

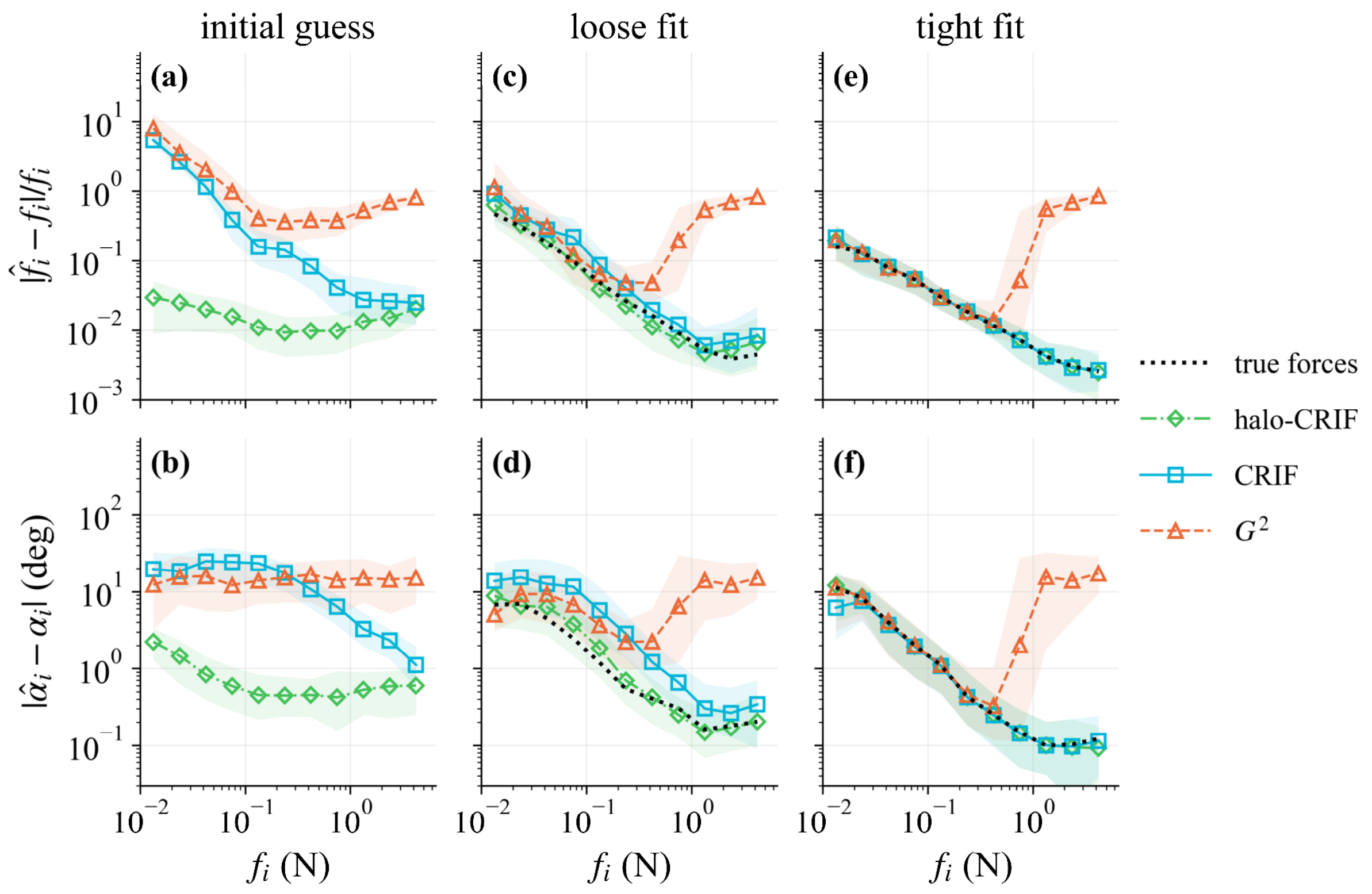

Fig. 8 Errors of initial guesses and fitted forces on 500 synthetic disks with halo. The top row shows the relative error in force $|\hat{f}_i - f_i|/f_i$ and the bottom row the angle error $|\hat{\alpha}_i - \alpha_i|$. (a, b) Initial guesses from halo-CRIF, trained on disks with residual stress; CRIF, trained on disks without it; and $G^2$. (c, d) Loose and (e, f) tight fits from these guesses with the forward model that includes the residual stress. The dotted line is the fit that started from the true forces, which sets the accuracy floor. Lines are bin medians and bands show the spread from the 25th to 75th percentile.

First, we compare the error of the initial guess from $G^2$, CRIF, and halo-CRIF on haloed synthetic images. Figures 8(a) and 8(b) show the halo-CRIF guess consistently outperforming the $G^2$ guess as expected. The CRIF guess is confused by the halo and loses its accuracy at $f_i < 0.2$ N, giving error levels similar to $G^2$; forces larger than 0.5 N produce distinct fringe patterns outside of the halo, which CRIF recognizes increasingly well, so that its error declines and approaches that of halo-CRIF.

As before, fits started from the $G^2$ guess fail in the large $f_i$ regime in all cases [Fig. 8(c–f)]. Other than that, all fitted results are close to the true-force floor, whether they start from CRIF or halo-CRIF. This is because the halo is accounted for in the fitting, and any guess within the right neighborhood can thus find good convergence. Nevertheless, halo-CRIF retains a noticeable advantage in the loose fit [Fig. 8(c, d)], where it already reaches the accuracy floor while the CRIF start does not. The difference suggests that training the CNN with halos can accelerate convergence in the fitting stage, while the halo forward model in the fit is what ultimately yields accurate results.

For a fair comparison, all fits above use the forward model that includes the halo. The consequence of omitting $\boldsymbol{\sigma_s}$ from the fit is shown in Appendix D.

### *5.3 Experimental validation and the reciprocity of force pairs*

Finally, we present an experimental application where a mixture of ClearFlex50 disks, 10 mm thick with diameters of 12 and 15 mm, is sheared in a fixed-volume channel. These disks have been immersed in 70% glycerol aqueous solution for more than a year, and maintain a constant halo as shown in Fig. 7. We first show in Fig. 9 the comparison between the experimental image and the synthetic images that visualize the force inversion results starting from the $G^2$ and halo-CRIF guesses. Qualitatively, fits started from halo-CRIF reproduce some of the more complicated fringe patterns much better than fits started from the $G^2$ guess, as highlighted by the orange circles.

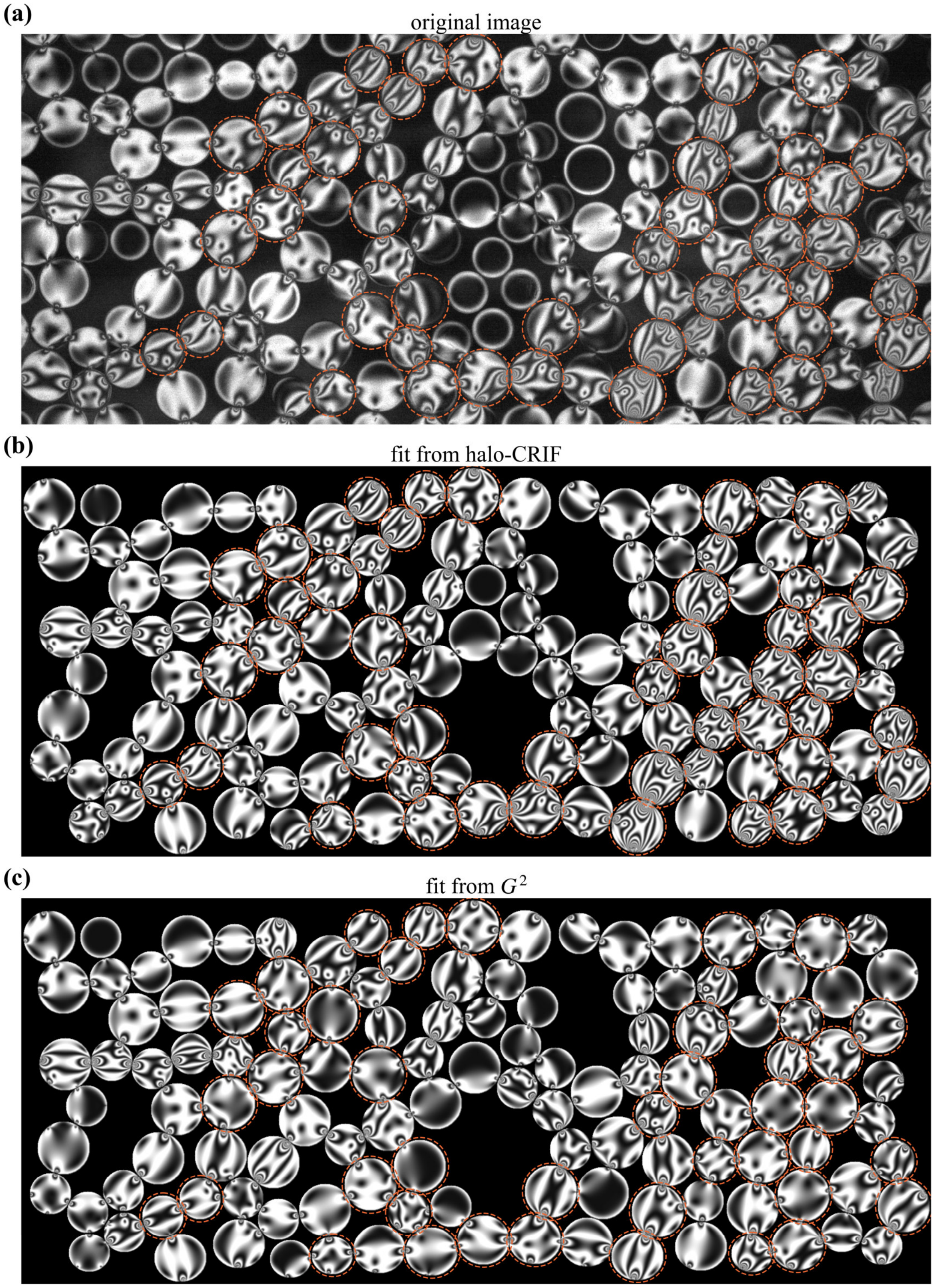

Fig. 9 A strongly loaded sheared packing. (a) Captured experimental image. (b) Images rendered from the fitted contact forces when the fit starts from the halo-CRIF guess, and (c) the $G^2$ guess. Both fits use the halo forward model and the tight-fit settings. Disks that were not fitted (fewer than two contacts, or cut by the image edge) are left black. Dashed orange circles mark the 42 of the 67 disks carrying at least one contact force above 1 N whose final loss from the $G^2$ start exceeds that from the halo-CRIF start by more than 30%. Note that the original image (a) is presented without post processing, so that its contrast appears lower than the two fits (b, c) that have been rescaled to the full bit depth (0 to 255).

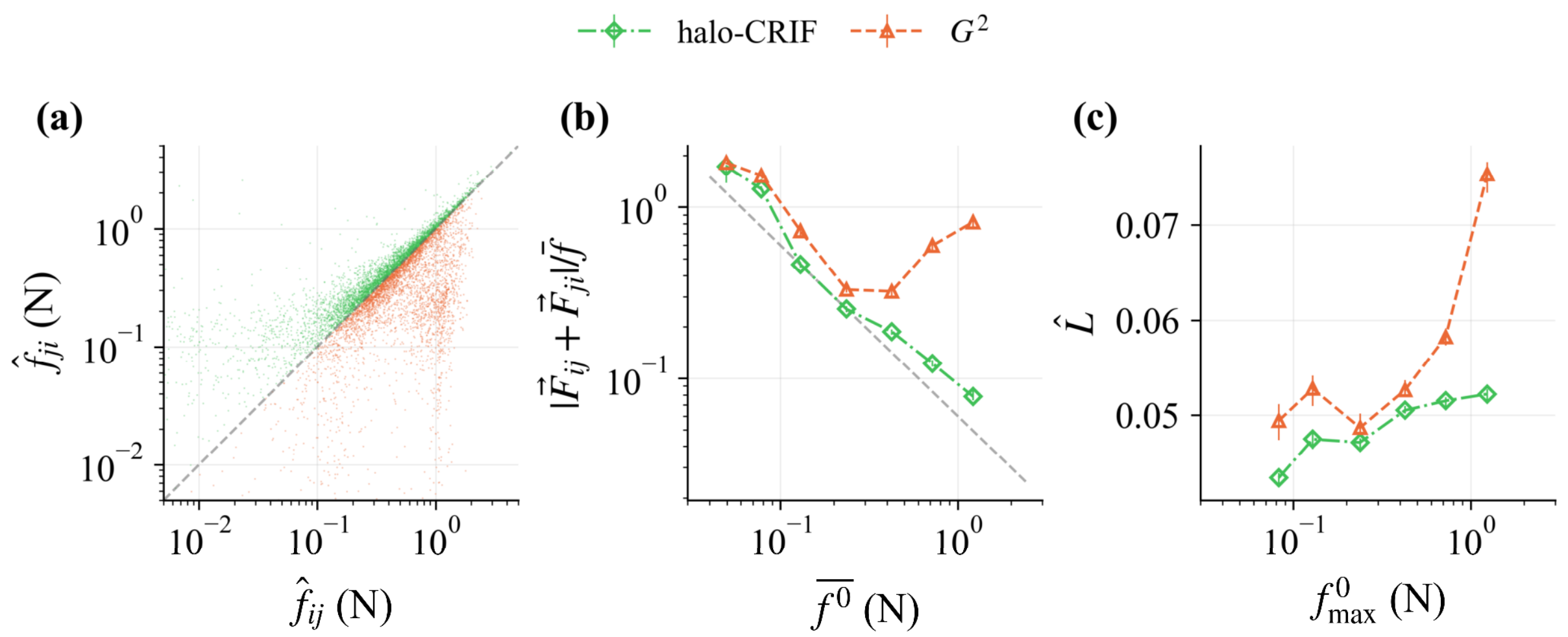

Fig. 10 Consistency of the fits on the packing, started from halo-CRIF (green) and from G² (orange). (a) The two independently fitted estimates of each contact force, $\hat{f}_{ij}$ from disk $i$ and $\hat{f}_{ji}$ from disk $j$, for all 4,334 contact pairs; the plot is symmetric about the diagonal, so halo-CRIF pairs are shown above it and G² pairs below. (b) Relative force vector mismatch $|\overrightarrow{F_{ij}}+\overrightarrow{F_{ji}}|/\bar{f}$. $\overline{f^0}$ on the horizontal axis represents each pair's mean halo-CRIF guess. Dashed line indicates $\propto f^{-1}$. (c) Final loss of each disk. Data points in (b,c) are bin medians.

To quantify the comparison, we fitted 24 frames with a total of 4,334 contact pairs. Since each disk is fitted independently, every contact force magnitude is fitted twice as $\hat{f}_{ij}$ and $\hat{f}_{ji}$. We check the reciprocity, i.e. the consistency with Newton's third law, by examining the difference $|\hat{f}_{ij}-\hat{f}_{ji}|$ over all contacts, which also provides an error estimate. Figure 10(a) plots the scatter of these two values, showing a growing relative mismatch at small forces and the failure of fits started from the $G^2$ guess for forces larger than 1 N. We calculate the relative force-vector mismatch $|\overrightarrow{F_{ij}}+\overrightarrow{F_{ji}}|/\bar{f}$, where $\bar{f}=(\hat{f}_{ij}+\hat{f}_{ji})\ /2$ is the pair-averaged magnitude. Since there is no true value as in the previous plots, we plot in Fig. 10(b) this mismatch against the pair-averaged halo-CRIF guess, $\overline{f^0}=(f^0_{ij}+f^0_{ji})/2$, as it is shown to provide relatively accurate guesses. The distinction between the two guesses becomes evident for $\overline{f^0}>0.3$ N, similar to the threshold in Figs. 6 and 8. Again, the relative vector mismatch of the halo-CRIF start shows a slope close to $-1$, indicating an absolute error that stays roughly constant at roughly 0.1 N regardless of force magnitude. Finally, we recorded the final fitting loss $\hat{L}$ of each disk, and plot it against the maximum initial guess by halo-CRIF on that disk, $f^0_{\max}$. The $G^2$ guess results in a much higher loss when the disk is subject to large forces, which implies that these fits end in wrong local minima.

## 6. Discussion

Across all scenarios, the main advantage of the CRIF guess is that it stays accurate above $f_i \approx 1$ N, independent of $Z$, which is where fits started from $G^2$ break down. In our case, this force corresponds to $N_{fr} \approx 1$, the point at which a full fringe forms across the disk. Below it, a start that is around 30% off, as the $G^2$ guess typically is, still lies in the basin of the correct solution. Beyond it, the high-fringe order image creates a complicated loss landscape: the $G^2$ guess can often be in the wrong basin so that the fit converges incorrectly. On the other hand, CRIF is trained on the fringe pattern at each contact itself and keeps its error at about 1%, so the fit converges up to the largest forces in our training set ($N_{fr} \approx 6$). Since stresses in granular packings are generally inhomogeneous and concentrate on an anisotropic strong network[3, 5, 31], a fraction of the particles often carry forces much larger than the mean on more than two contacts (see Fig. 9 for an example). The ability to resolve large forces on $Z > 2$ disks could thus significantly improve the robustness of photoelastic analysis on granular experiments.

In some cases, finding $f_i$ and $\alpha_i$ on each contact to the optimized precision is not strictly necessary. For example, many past experiments used only the $G^2$-inferred total force $f_{tot}$ on each particle[9, 17–19]. In these cases, we suggest that the fitting step can be optionally omitted, and the guess from CRIF could directly serve as a good estimate. This is demonstrated in Fig. 11, where we calculate the boundary force in the y direction (the vertical direction in Fig. 9) of a shear run using halo-CRIF estimated forces, in comparison with direct measurements from a boundary load cell. The two agree reasonably well. In the case of our data, the runtime for the halo-CRIF guess is around 30% of the loose fit, or 1% of the tight fit. Skipping the fitting can thus save considerable computational cost if the level of precision is acceptable.

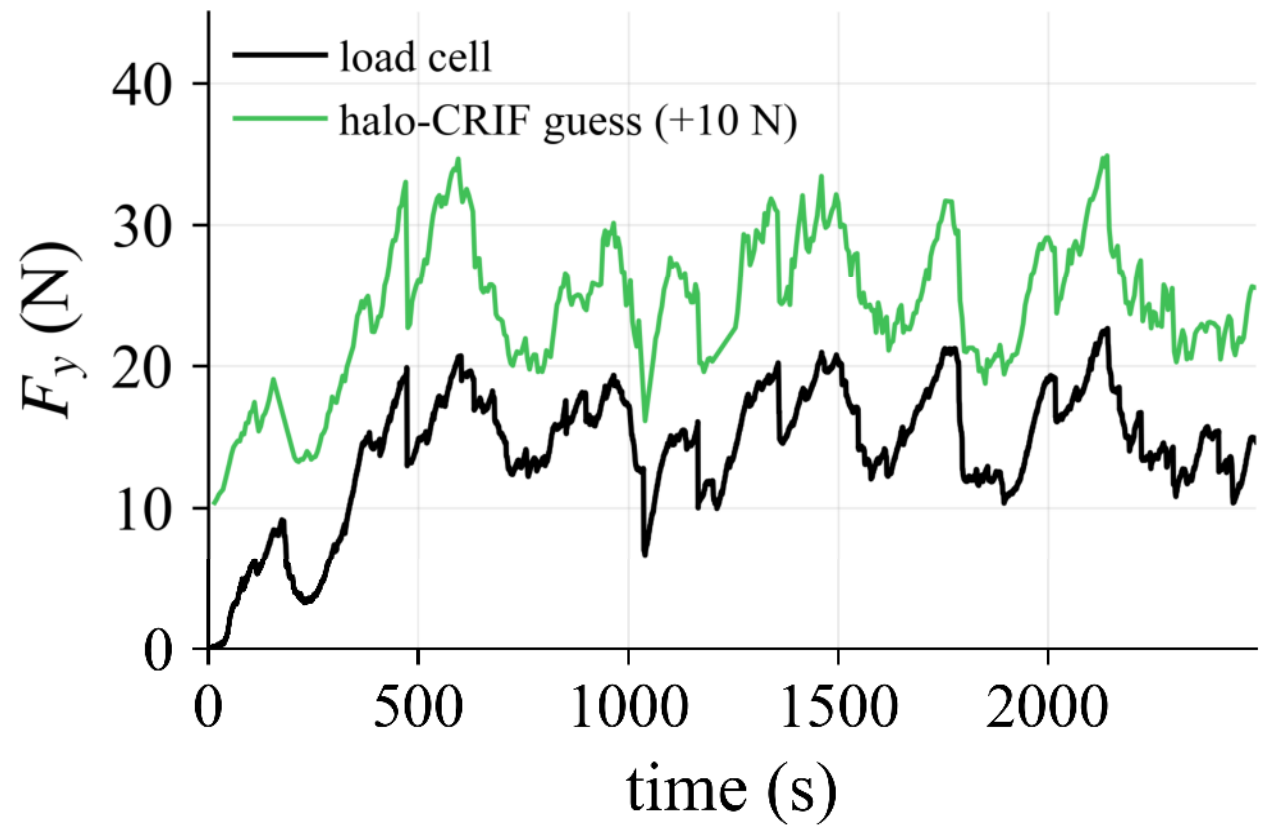

Fig. 11 Boundary force $F_y$ during steady shear at shear rate $1.2 \times 10^{-3}$ s$^{-1}$, measured by the load cell (black) and obtained from the halo-CRIF guesses alone, without fitting (green). The guess is shifted up by 10 N for clarity.

We emphasize that, like all neural networks, the capability of CRIF is defined largely by its training data. We demonstrated the potential applications in Sec. 5 by introducing a common residual halo. In principle, the same approach applies to any forward model that can generate synthetic training images, analytic or numerical. Polygons and other non-circular grains, for example, have no general closed-form stress solution, so every forward evaluation requires a slow finite-element solve, which is impractical inside an iterative fit. A network trained once on finite-element renderings could potentially predict contact forces for such systems.

For readers who wish to train their own models, we note that the model architecture and training hyperparameters used herein (Appendix C) are not optimized for this specific task, and are chosen arbitrarily to represent generic settings. A lighter model could potentially deliver the same accuracy with better efficiency. In addition, care should be taken when choosing the image resolution as well as modification parameters in the data generation process, such that the synthetic images best mimic the experimental ones. Ideally, one could also fine-tune the trained network with a smaller set of calibration images with known loads on real disks[21], which would optimize the network for real images. This is not included in this study due to the experimental challenge of imposing multiple accurate point loads on a disk, while also generating a statistically meaningful number of configurations. The forward model, dataset generator, training code and trained networks are publicly available [25].

## 7. Conclusion

We have introduced CRIF, which uses a convolutional neural network to infer the force at each contact of a photoelastic disk from a small region around that contact. Trained only on synthetic images rendered with the forward model, it predicts the contact forces of unseen test disks with a median error of about 1% in magnitude and 0.5° in direction, before any fit, over nearly three decades of force and for up to six contacts per disk. As an initial guess for a PeGS-style fit, it brings the fitted forces to the accuracy floor set by the true forces across the whole range, including the strongly loaded disks where the $G^2$ guess leads the fit into a wrong local minimum.

Because the network learns from the forward model, it can adapt to changes in the physics of the disks. For disks with a residual-stress halo, including the halo in the forward model of the fit is what recovers accurate forces, and a network retrained on haloed images reaches that accuracy faster. On a sheared packing of immersed disks, fits started from halo-CRIF are consistent with Newton's third law between the two disks of each contact, where fits started from the $G^2$ guess fail for strong contacts.

CRIF does not necessarily replace the physics-based PeGS fit; it gives the fit a better place to start. The approach can be extended to other materials, particle shapes and imperfections by rendering a corresponding training set, which makes quantitative force inversion possible in regimes, such as large forces, many contacts or immersed particles, that have so far been difficult to reach.

## Code availability

The code used in this study is openly available on GitHub at https://github.com/jctsaigroup/CRIF_for_photoelasticity under the MIT license. It includes the photoelastic forward model, the synthetic dataset generator, the training and inference code, and the two trained networks (clean and halo). The version used for this paper (v1.0.1) is archived on Zenodo at https://doi.org/10.5281/zenodo.24358576 [25].

## Appendix A: Stress field of a contact

In the polar coordinates $(r_i, \theta_i)$ centered at the $i$th contact and measured from the direction of the force (Fig. 1), the Flamant solution for a point force $f_i$ is purely radial[26]:

$$\sigma_{rr,i}^0(r_i, \theta_i) = -\frac{2f_i\cos(\theta_i)}{\pi r_i} \qquad Eq. A1$$

with all other components zero. In Eq. 2, $\boldsymbol{T}(\gamma_i)$ is the two-dimensional rotation matrix associated with the orientation angle $\gamma_i = \theta_i - \alpha_i - \phi_i$ of the $i$-th contact, which brings $\boldsymbol{\sigma}_i^0$ into the Cartesian frame of the disk.

## Appendix B: Synthetic force sets and rendering

The contact angles $\phi_i$ are first distributed randomly on the circle, conditioned on a minimum angular separation of $\theta_{min} = 53°$. This lower bound is looser than the 60° required by a monodisperse packing, and is chosen to provide generality for a bidisperse system with a size ratio of 1.25. The total force magnitude $f_{tot}$ is then drawn log-uniformly between 0.02 and 4 N. The normal components are $f_i^n = \varepsilon_i f_{tot}$, with $\ln\varepsilon_i$ normally distributed with zero mean and standard deviation 0.5. The tangential components are $f_i^t = f_i^n \tan\alpha_i$, with $\alpha_i$ drawn from a uniform distribution within $[-40°, 40°]$. Note that the resulting sum of forces does not equal $f_{tot}$, which only serves to set the overall scale.

Equilibrium is enforced by fixing contact position $\{\phi_i\}$ and adjusting $\{f_i, \alpha_i\}$. We can express the force vectors as $\mathbf{x}$, a vector with $2Z$ components. The two net forces and net torque can similarly be written as $\mathbf{Cx}$, where $\mathbf{C}$ is a $3 \times 2Z$ matrix that only depends on $\{\phi_i\}$. We find the correction $\Delta\mathbf{x}$ that satisfies $\mathbf{C}(\mathbf{x} + \Delta\mathbf{x}) = 0$ and minimizes $|\Delta\mathbf{x}|^2$. With a Lagrange multiplier, the solution is $\Delta\mathbf{x} = \mathbf{C}^{\mathrm{T}}(\mathbf{CC}^{T})^{-1}\mathbf{Cx}$, which is then applied to impose equilibrium. Finally, we reject force sets after the equilibrium correction if (1) any normal component falls below $0.02f_{tot}$; (2) $f_i^t > f_i^n$ at any contact; (3) any force exceeds 6.4 N. These filters regularize our dataset to a well-defined range. The resulting forces span 0.01–6.4 N, with nearly uniform coverage in $\ln f$, and force angles up to 45° (Fig. 2(a, b)).

Rendering is done at four times the final resolution and averaged over $4 \times 4$ blocks, so that fringes finer than a pixel average to gray, as on a camera sensor. We then apply a random gain and offset, Gaussian blur, additive Gaussian noise, and finally quantization to 8 bits, to the image with parameters drawn independently for each image from the ranges in Table A1.

In practice, the gain and offset of the experimental image are a fixed set of values that can ideally be measured and input into the fitting process. The Gaussian noise and blur remain as unavoidable imperfections of the camera, and the noise is amplified by a factor of $1/g$ in the process. We do not normalize each disk by its own maximum since weakly loaded disks never reach full brightness.

**TABLE A1.** Image modification parameters.

| | **value** |
|---|---|
| Gaussian blur (standard deviation) | 0.5–1.5 pixels |
| Gain $g$ | 0.6–1.0 |
| Offset $b$ | 0–0.05 |
| Gaussian noise (standard deviation) | 0.005–0.03 |

## Appendix C: Network training

The classifier of the pretrained ResNet-18 is replaced by a fully connected layer of 256 units (ReLU activation, dropout 0.2) followed by two outputs. The gray-level crop is replicated over the three input channels and normalized with the ImageNet statistics. The only augmentation is a left–right flip, which reverses the sign of $\alpha$. The two targets are:

$$y_1 = \frac{\ln(\frac{f}{R}) - \mu_f}{s_f}, \qquad y_2 = \frac{\alpha}{\pi/4} \qquad Eq. A2$$

, where $\mu_f$ and $s_f$ are the mean and standard deviation of $\ln(f/R)$ for normalization. We use $f/R$ , which has units of a planar stress, as the target, since it determines the fringe pattern in Eq. 2. The model is thus trained to be applicable to disks of different $R$ but the same $f_\sigma$. The logarithm gives relative errors equal weight across the nearly three decades of force. The loss function is the sum of the mean squared errors of $y_1$ and $y_2$. Training proceeds in two phases with the AdamW optimizer. In a warm-up phase the pretrained layers are frozen and only the new head is trained; in the fine-tuning phase all layers are trained, and the learning rate is halved whenever the validation loss stops improving (Table A2). We keep the model with the lowest validation loss.

**TABLE A2.** Training parameters for the CRIF network

| | |
|---|---|
| batch size | 64 |
| Fully connected layer size | 256 |
| Dropout | 0.2 |
| warm-up (head only) | 10 epochs, learning rate $10^{-3}$ |
| fine-tuning (all layers) | learning rate $10^{-4}$, weight decay $10^{-4}$ |
| | halved after 4 epochs without 0.5% improvement |
| | stop below $10^{-6}$ or after 12 such epochs |

Figure A1 shows how the accuracy of the guess depends on the size of the training set. Diminishing returns are apparent beyond $2 \times 10^4$ contacts. For ideal convergence, at least $10^4$ training contacts are recommended. A second training on the full set with a different random seed gives similar median force error of 1.1%, which validates the consistency across different random initializations.

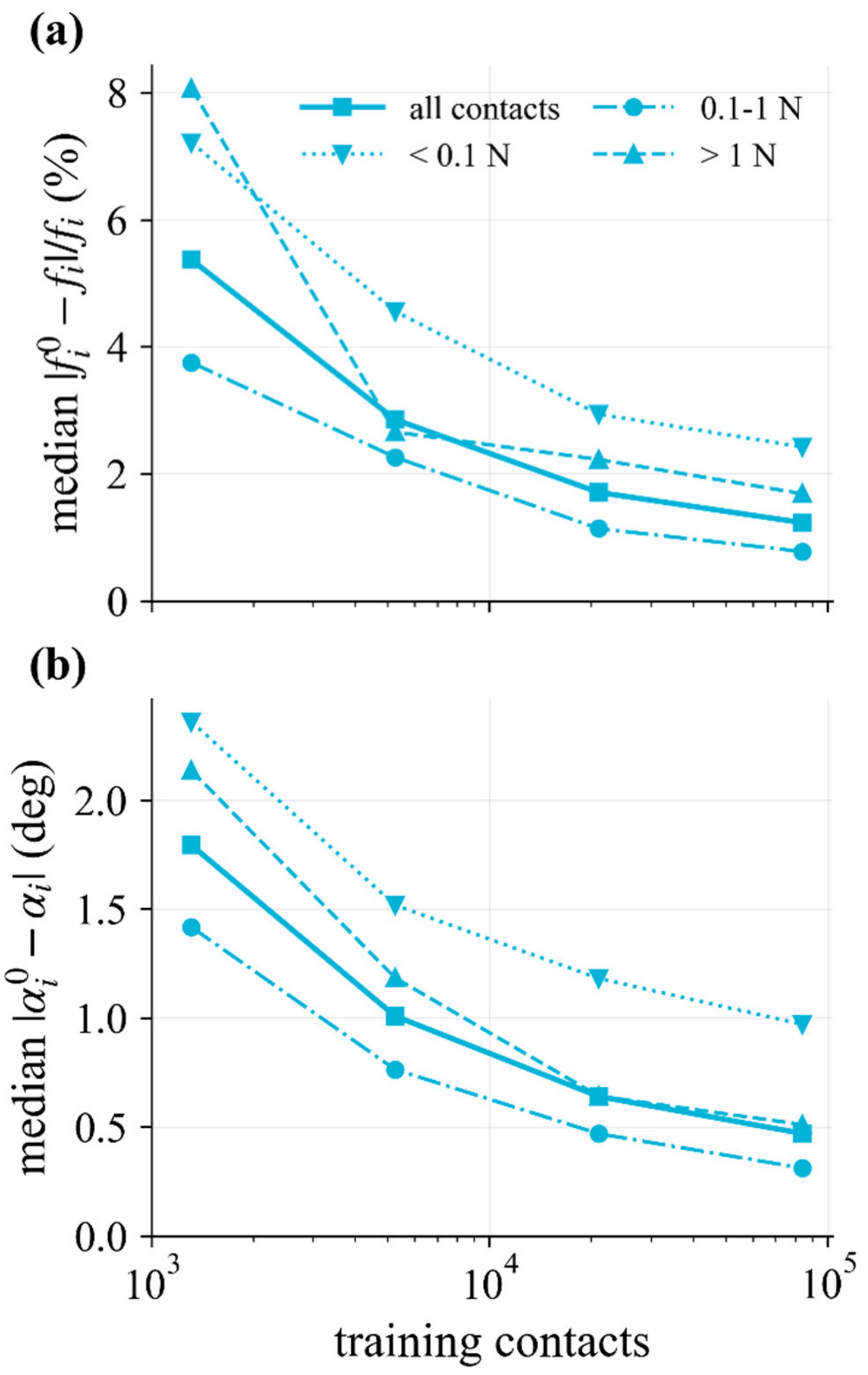

Fig. A1 Median relative force error (a) and median angle error (b) of the CRIF initial guess on the synthetic test set, against the number of training contacts, for all contacts and for contact forces below 0.1 N, between 0.1 and 1 N, and above 1 N. Each network was trained with the same recipe on a random subset of the training set.

## Appendix D: Fitting without the residual stress

To show the consequence of omitting the residual stress from the fit, we fit the 500 haloed test disks of Sec. 5.2 with the original forward model (Eq. 2), starting from the $G^2$ and CRIF guesses with the tight-fit settings (Fig. A2). Here CRIF is the network trained without residual stress. Fits started from the CRIF guess still reach the floor of this mismatched model, but the floor itself is poor for small forces. Errors in $f_i$ are larger than 10% for forces smaller than 0.3 N, and the angles are off by up to 80 degrees at the smallest forces, since the optimizer mistakenly makes the forces almost completely tangential so that fringes run along the edge of the disk. These results highlight the necessity of incorporating appropriate residual stresses in the forward model when they are present.

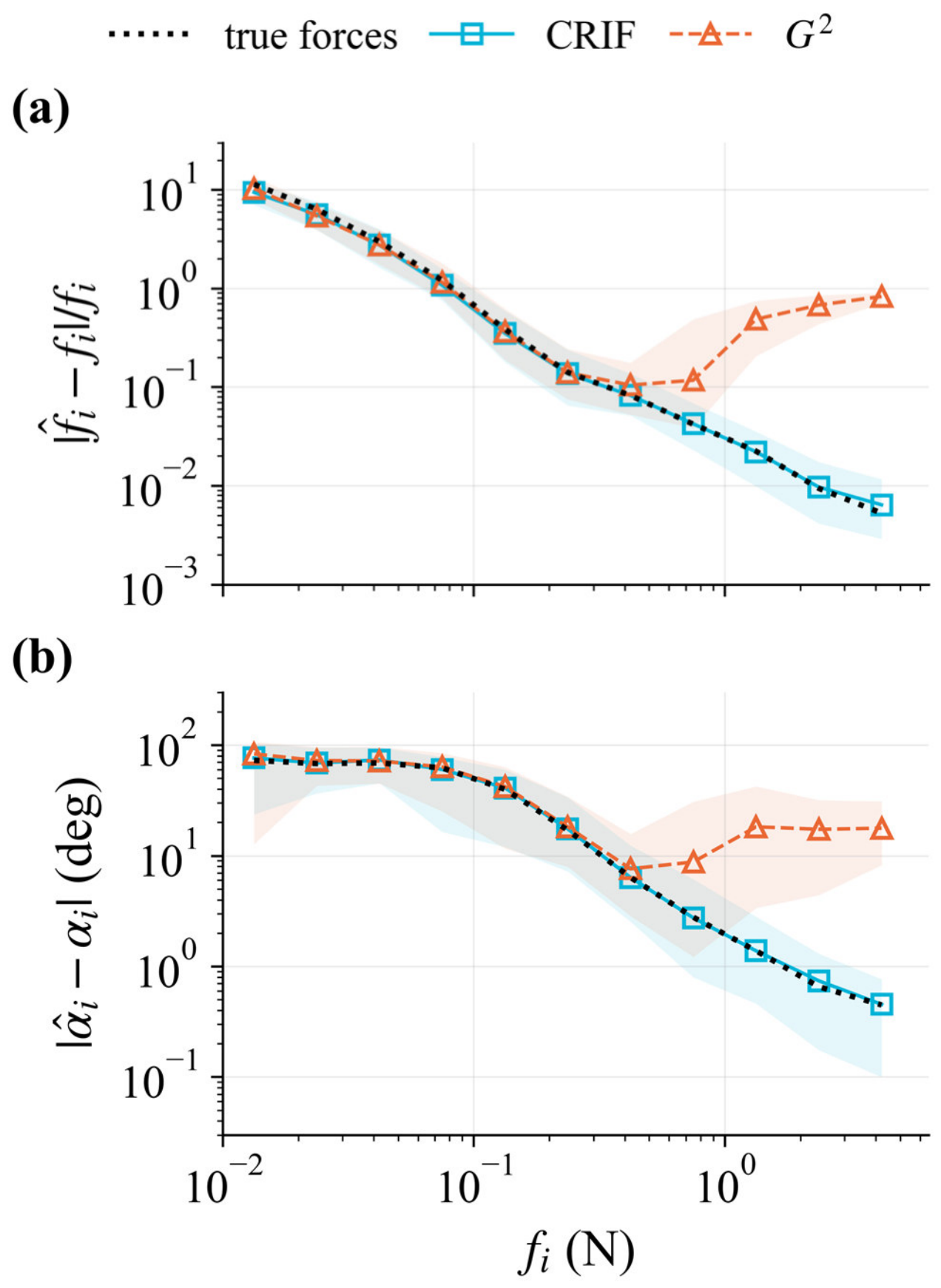

Fig. A2 Tight fits on the 500 synthetic disks with halo when the forward model omits the residual stress: (a) relative force error and (b) angle error against the true force. The fits start from the CRIF guess (network trained without residual stress), the G² guess, and the true forces (dotted line), which sets the accuracy floor of this forward model. Lines are bin medians and bands show the spread from the 25th to 75th percentile.